\documentclass[aps,pre,twocolumn,showpacs]{revtex4}
\usepackage{epsfig}
\usepackage{times}
\usepackage{amsmath}

\begin{document}

\title{Phase diagrams for three-strategy evolutionary prisoner's dilemma games on regular graphs}

\author{Attila Szolnoki$^1$, Matja{\v z} Perc$^2$, and Gy{\"o}rgy Szab{\'o}$^1$}
\affiliation
{$^1$Research Institute for Technical Physics and Materials Science,
P.O. Box 49, H-1525 Budapest, Hungary \\
$^2$Department of Physics, Faculty of Natural Sciences and Mathematics, University of \\ Maribor, Koro{\v s}ka cesta 160, SI-2000 Maribor, Slovenia}

\begin{abstract}
Evolutionary prisoner's dilemma games are studied with players located on square lattice and random regular graphs defining four neighbors for each one. The players follow one of the three strategies: tit-for-tat, unconditional cooperation, and defection. The simplified payoff matrix is characterized by two parameters: the temptation $b$ to choose defection, and the cost $c$ of inspection reducing the income of tit-for-tat. The strategy imitation from one of the neighbors is controlled by pairwise comparison at a fixed level of noise. Using Monte Carlo simulations and the extended versions of pair approximation we have evaluated the $b-c$ phase diagrams indicating a rich plethora of phase transitions between stationary coexistence, absorbing and oscillatory states, including continuous and discontinuous phase transitions. By reasonable costs the tit-for-tat strategy prevents extinction of cooperators across the whole span of $b$ values determining the prisoner's dilemma game, irrespective of the interaction graph structure. We also demonstrate that the system can exhibit a repetitive succession of oscillatory and stationary states upon changing a single payoff value, which highlights the remarkable sensitivity of cyclical interactions on parameters that define the strength of dominance.
\end{abstract}

\maketitle

\section{Introduction}
Cyclical interactions underlie the mating strategies of side-blotched lizards \cite{sinervo_n96}, the competition between different strains of \textit{Escherichia coli} \cite{kerr_n02}, and the overgrowth of marine sessile organisms \cite{burrows_mep98}. The paradigmatic example, however, is the children's rock-paper-scissors game \cite{hofbauer_98}, which despite its simplicity still maintains notable scientific allure \cite{frean_prsb01, szolnoki_pre04, reichenbach_prl07, efimov_pre08}. Models of cyclical interactions are used to tackle fundamental problems in theoretical biology and ecology, as well as social and economic systems, whereby sustainability of biodiversity \cite{czaran_pnas02} and the evolution of cooperation \cite{hauert_s02} appear to be the two most prominent topics. Arguably thus, cyclical interactions are fascinating and powerful examples of evolutionary processes, able to provide insights into the intriguing mechanisms of Darwinian selection \cite{maynard_n73} as well as structural complexity and prebiotic evolution \cite{rasmussen_s04}.

Evolutionary game theory \cite{maynardsmith_82} provides a competent framework in which to investigate the success of different strategies or species in well-mixed as well as structured populations \cite{nowak_06}. In this context the rock-paper-scissors and related games have been investigated intensely \cite{szabo_jpa04,szabo_pr07, traulsen_09, szollosi_pre08}, and particularly the effects of mobility \cite{reichenbach_n07} and stochasticity \cite{reichenbach_pre06} recently received substantial coverage in view of potential maintenance of biodiversity. The evolution of cooperation within models incorporating a closed loop of dominance has been addressed within public goods games \cite{hauert_s02, szabo_prl02}, where it has been shown that volunteering leads to rock-paper-scissors dynamics between the participating strategies \cite{semmann_n03}. Similar observations have been made for the prisoner's dilemma game as well. In general, the modifications brought about by strategic complexity have been found favorable for the sustenance of cooperation, either in terms of stationary or oscillatory states \cite{peltomaki_pre08b}.

An interesting extension of the prisoner's dilemma game has been proposed by Imhof \textit{et al.} \cite{imhof_pnas05}, who besides the unconditional cooperation and defection introduced the well-known tit-for-tat strategy as a third type using mutation-selection dynamics in well-mixed population. The tit-for-tat strategy has proven most successful for the iterated prisoner's dilemma game \cite{axelrod_84} even within the set of stochastic reactive strategies \cite{nowak_n92b}. Similar to equivalent retaliation or reciprocity, the virtue of the tit-for-tat strategy is to follow the opponents previous action, albeit initially to always cooperate. Only few concepts were thus far able to challenge the success of this fairly simple yet effective strategy \cite{nowak_n93, imhof_jtb07}. In Ref.~\cite{imhof_pnas05} authors reported that the introduction of the tit-for-tat strategy to the prisoner's dilemma game via mutation-selection dynamics in finite \cite{nowak_n04} well-mixed populations leads to a natural selection of reciprocity, thus sustaining cooperation where otherwise defection would dominate completely.

Here we extend the subject by studying the evolutionary prisoner's dilemma game supplemented by the tit-for-tat strategy in structured populations, showing that cooperation can survive even in the absence of mutation. The prisoner's dilemma game is a paradigmatic example of a social dilemma \cite{macy_pnas02}, which even with the addition of the tit-for-tat as a third strategy, features defection as the strict Nash equilibrium \cite{imhof_pnas05}. Since spatial structure may maintain cooperative behavior in the prisoner's dilemma game, as has been shown in the seminal work of Nowak and May \cite{nowak_n92}, it is also of interest to investigate the evolutionary outcome of the three-strategy version, entailing cooperation, defection and tit-for-tat as the possible strategies. Since the classical mean-field approximation assumes well-mixed populations, we use the $k$-site cluster dynamical mean-field approximation to capture the essential role of different geometries of the connectivity structure. The dynamical mean-field approximation technique proved to be a powerful tool for the determination of phase diagrams in several non-equilibrium systems \cite{dickman_pre01, szabo_jpa04, szabo_pr07}. Indeed, we show that for the considered evolutionary game the phase diagrams obtained subsequently via Monte Carlo simulations are in qualitative agreement with the $k$-site cluster dynamical mean-field approximation, evidencing a rich dynamical behavior depending on the payoff parameters. The payoff parameters determine the strength of cyclical interactions between the three strategies and, depending also on the interaction graph, crucially influence the outcome of the game, both in terms of strategy abundance as well as the resulting dynamics.

The remainder of this paper is organized as follows. In the next section we describe the three-strategy evolutionary game and present the predictions of the dynamical mean-field approach. These results are compared with the output of Monte Carlo simulations in Section III, whereas in the last Section we summarize our findings and discuss their potential implications.

\section{Game definition and dynamical mean-field approximations}

We extend the two-strategy (unconditional cooperation C and defection D) spatial prisoner's dilemma game (for a survey see \cite{nowak_06,szabo_pr07}) with allowing the players to use the tit-for-tat (in short T) strategy, too. Within the framework of two-strategy game we adopt the parametrization suggested by Nowak and May \cite{nowak_n92}, i.e. the prisoner's dilemma is characterized by the temptation $b$, reward $1$, and both punishment as well as the suckers payoff equaling $0$, whereby $1 < b \leq 2$ ensures a proper payoff ranking. We should stress, however, that our observations are not restricted to the so-called weak prisoner's dilemma limit but remain fully valid also if the rank of payoff elements fulfills the game definition strictly.

We introduce the tit-for-tat strategy so that we neglect the payoff reduction of T arising from the first encounter with a defector \cite{szabo_pre00a}. This simplification, still retaining the essence of the payoff matrix, is valid in the case players imitate the neighbor's strategy rarely in comparison with the frequency of games they play. In many previous models \cite{doebeli_pnas97,imhof_jtb07} the payoff reduction of T arising from the first cooperation is considered quantitatively and caused technical difficulties in the investigations. As the application of T strategy requires continuous inspection and recording the neighbors' previous step, the pure income of T players is reduced by the cost of monitoring. Accordingly, the reward (against C or T) is $1-c$, whereas a defector gets $0$ if facing a T player who then obtains $-c$. The payoff elements are summarized in Table~\ref{payoff}.

\begin{center}
\begin{table}
\caption{Payoff matrix of the studied evolutionary game. The three strategies are: cooperation (C), defection (D), and tit-for-tat (T). $b$ is temptation to defect, $c$ is the cost of monitoring other's strategy.}
\begin{tabular}{llll}
& \vline \,\, C & \,\,\,D & \,\,\,\,\,\,\,\,\,T \\
\hline
C & \vline \,\,\,$1$ &\,\, $0$ &\,\,\,\,\,\,\,\, $1$ \\
D & \vline \,\,\,$b$ &\,\, $0$ &\,\,\,\,\,\,\,\, $0$ \\
T & \vline \,\,\,$1-c$ & \,\,$-c$ &\,\,\,\,\,\,\,\, $1-c$  \\
\end{tabular}
\label{payoff}
\end{table}
\end{center}

Note that the presence of tit-for-tat players introduces a cyclic dominance-type relation among strategies in the spatial model, which is absent in the well-mixed case. More precisely, a group of neighboring players adopting the tit-for-tat strategy can spread in the sea of D's because they support each other as cooperators yet do not allow defectors to exploit them. On the other hand, even a single cooperator can invade tit-for-tat players because their cooperation is burdened by the monitoring cost. Furthermore, above a threshold value of $b$ defectors conquer cooperators in the absence of T. The mentioned relations introduce a sort of cyclic dominance between the three strategies (C $\rightarrow$ T $\rightarrow$ D $\rightarrow$ C). More importantly, the intensity of dominance can be tuned effectively via alterations of $b$ and $c$. Accordingly, the present study may strengthen the link between evolutionary game models and other systems of cyclic dominance relevant to biology and social sciences \cite{cyc}.

The game is staged on a $L \times L$ square lattice with nearest neighbor interactions and periodic boundary conditions or a random regular graph with an identical degree, whereon initially each player on site $i$ is designated either as a cooperator ($s_i =$~C), defector (D) or a tit-for-tat player (T) with equal probability. Irrespective of the interaction graph, a randomly selected player $i$ acquires its payoff $P_i$ by playing the game with its four neighbors. Next, one randomly chosen neighbor, denoted by $j$, also acquires its payoff $P_j$ by playing the game with its four neighbors. Last, player $i$ tries to enforce its strategy $s_i$ on player $j$ in accordance with the probability
\begin{equation}
W(s_i \rightarrow s_j)=\frac{1}{1+\exp[(P_j-P_i)/K]},
\label{eq:prob}
\end{equation}
where $K$ denotes the amplitude of noise \cite{szabo_pre98}. For simplicity we keep the value $K=0.1$ fixed throughout this study, but note that the qualitative features remain unchanged for different $K$.

\begin{figure}
\centerline{\epsfig{file=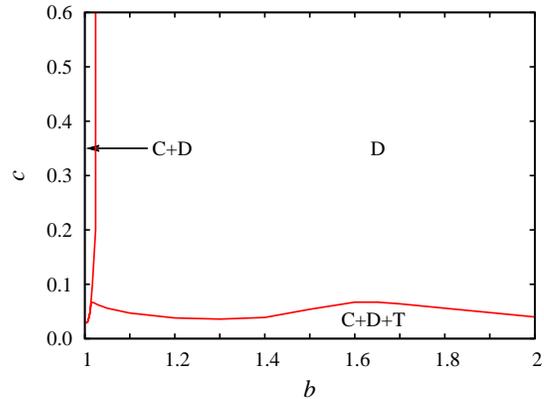,width=8cm}}
\caption{(color online) Full $b-c$ phase diagram for the three-strategy evolutionary game on the square lattice obtained via the $4$-site (square) cluster dynamical mean-field approximation. Red lines mark the border between stationary mixed states C+D and C+D+T, as well as absorbing D ($\rho_{\rm D}=1$) states.}
\label{phd_4p}
\end{figure}

\begin{figure}
\centerline{\epsfig{file=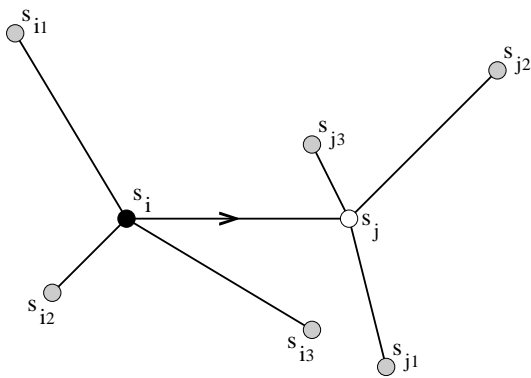,width=7.4cm}}
\caption{Schematic presentation of the evolution of 2-site cluster probabilities on a random regular graph. The central players are $s_i$ (black circle) and $s_j$ (white circle) with four neighbors each (gray circles and the corresponding central player). Black player passes strategy to the white player, as denoted by the arrow.}
\label{2p_dmf}
\end{figure}

We first present phase diagrams obtained via $k$-site cluster dynamical mean-field approximations. The first level that can approximate the square lattice structure (containing short loops) correctly is the $4$-site level, where the master equations for probabilities of configurations on a $2 \times 2$ cluster need to be derived. The details of this approach can be found in Ref.~\cite{dickman_pra90}, especially for three-strategy system in Ref.~\cite{szabo_jpa04}. The predictions of the $4$-site level approximation for the present model are summarized in Fig.~\ref{phd_4p}. It can be observed that for low values of $c$ a mixed stationary state of all three strategies exists across the whole span of $b$. For very low $b$ and $c>0.028$, however, the T strategy dies out, yielding a mixed C+D state, as marked by the arrow. In this region the system is equivalent to the traditional two-strategy prisoner's dilemma game. In the remainder of the phase diagram absorbing D states prevail. In essence, it can be concluded that the addition of the tit-for-tat strategy to the evolutionary prisoner's dilemma game on the square lattice maintains cooperation across the whole span of $b$ in the form of a stationary mixed C+D+T state if only the monitoring cost $c$ is sufficiently low. It should be noted, however, that the non-monotonous $b$ dependence of the transition line separating C+D+T and D phases is just an artifact originating from the order of approximation.

As expected, the lack of `shortcuts' (or the absence of the small-world property) prevents the emergence of synchronized strategy invasions. Thus, the mixed C+D+T phase depicted in Fig.~\ref{phd_4p} has the properties of a stationary state in that the permanent spreading of invasion fronts between the three strategies maintains their densities constant. In other words, at low values of the cost $c$, the self-organizing pattern of strategy distribution game on the square lattice has the same morphology as was observed previously for spatial rock-paper-scissors game \cite{szabo_pre99, szolnoki_pre04, reichenbach_prl08, peltomaki_jsm09}.

Based on the established conceptual similarity with the rock-paper-scissors model, significantly different behavior is expected if the square lattice interaction topology is replaced with the random graph. For the purposes of simplicity and comparability, we use random regular graphs ensuring the same degree $z=4$ for all players. Such a graph can be obtained by randomly rewiring the links of the original square lattice, as demonstrated in Ref.~\cite{szabo_jpa04}. In the absence of strong local correlations between neighbors of a given player, already the $2$-point level of dynamical mean-field approximation yields reliable predictions. At this level, the variables are denoted by $p_2(s_i,s_j)$, characterizing the configuration probability of a link connecting players with $s_i$ and $s_j$ strategies. By neglecting higher order correlations, the time derivative of $2$-point probabilities is given as a function of probabilities of such configurations. To illustrate these equations, the probability of a strategy adoption process between players $i$ and $j$ is given as
\begin{equation}
\frac{\displaystyle{p_2(s_i,s_j) \prod_{k,l=1}^3 p_2(s_i,s_{ik}) p_2(s_j,s_{jl})}}{[p_1(s_i) p_1(s_j)]^3} \, W(s_i \rightarrow s_j) \,,
\label{eq:adopt}
\end{equation}
where $p_1(s_i)$ and $p_1(s_j)$ denote the frequency of strategy $s_i$ and $s_j$, respectively, and the payoff difference dependent adoption rate $W$ is from Eq.~\ref{eq:prob}. The notation used for the description of the neighborhood of the $i-j$ link is illustrated in Fig.~\ref{2p_dmf}.

\begin{figure}
\centerline{\epsfig{file=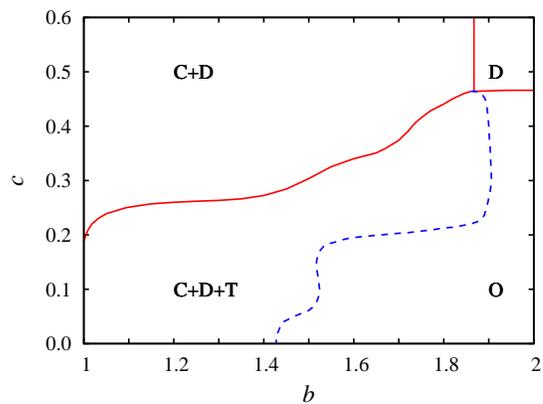,width=8cm}}
\caption{(color online) Full $b-c$ phase diagram for the three-strategy evolutionary game on the random regular graph obtained via the $2$-site cluster dynamical mean-field approximation. The phase diagram contains all the stationary states (mixed and pure) presented in Fig.~\ref{phd_4p}, as well as an additional region with oscillatory states denoted with O, which may be separated by a continuous second order (red line) or a discontinuous first order (dashed blue line) phase transition to absorbing D or stationary C+D+T phases, respectively.}
\label{phd_2p}
\end{figure}

\begin{figure}
\centerline{\epsfig{file=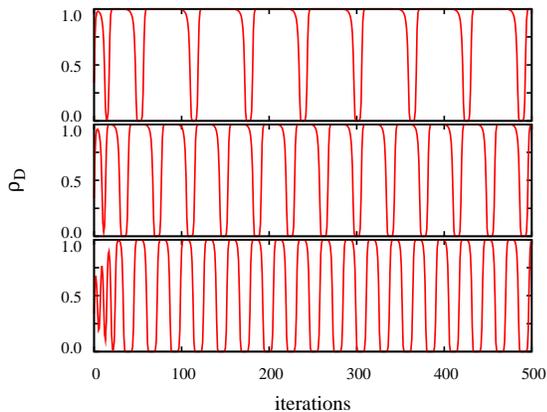,width=8cm}}
\caption{(color online) Time evolution of the density of defectors $\rho_{\rm D}$ for different values of $c$, as obtained via the $2$-site cluster dynamical mean-field approximation with $b=1.95$ being fixed. Values of $c$ are: $0.25$ (bottom panel), $0.35$ (middle panel), and $0.43$ (top panel). As $c$ is increased the system spends more and more time in the state where almost every player adopts the D strategy.}
\label{rare}
\end{figure}

The resulting phase diagram, presented in Fig.~\ref{phd_2p}, features, besides the stationary mixed (C+D and C+D+T) and pure (D) phases, also an oscillatory region (O), where all three strategies coexist in an oscillating manner. Interestingly, the transition from the stationary mixed C+D+T state to the oscillatory state is always discontinuous, denoted explicitly by the dashed blue line. We note that the starting amplitude of oscillatory states increases when $c$ is increased along the border separating C+D+T and O phases. However, the jump associated with this transition at $c \simeq 0.3$ becomes so sharp that the system instantly switches to the oscillatory state having maximal amplitude. Accordingly, the three strategies cyclically alternate almost complete dominance sequentially (C $\rightarrow$ T $\rightarrow$ D $\rightarrow$ C) in an oscillatory fashion. On the other hand, the transition from the oscillating O to the absorbing D phases is always continuous. The latter transition is illustrated explicitly in Fig.~\ref{rare}, where the time evolution of three different oscillating states is plotted, as obtained for three different values of $c$ by a fixed temptation $b=1.95$. At first glance, the oscillatory solutions vary only slightly in dependence on $c$, the most significant difference being that the duration of the almost complete dominance of D becomes increasingly longer as $c$ is increased. Stated differently, the departures to the C and T strategies become rearer, until eventually they are completely left out when the O-D transition line is crossed. It is worth noting that if plotted in the ternary diagram, all three oscillatory states depicted in Fig.~\ref{rare} would be identical, following closely the border of the triangle. Regarding the phase diagram, the high $b_c$ value separating C+D and D phases is again an artifact due to the low level of approximation, as we will show below. Next, it is thus of interest to test the accuracy of predictions obtained via the $k$-site cluster dynamical mean-field approximations with Monte Carlo simulations.

\section{Monte Carlo simulations}

Monte Carlo results presented below were obtained on populations comprising $400 \times 400$ to $5000 \times 5000$ individuals, whereby the fractions of the three strategies $\rho_s; s \in$~(C,D,T) were determined after discarding an appropriate relaxation time. The duration of this time depends significantly on the phase towards which the system is evolving. For example, at small $b$ and $c$ values $10^3$ Monte Carlo steps (MCS) are sufficient to reach the stationary state. On the other hand, more than $2 \cdot 10^4$ MCS need to be discarded at the border of C+D+T and O phases. Furthermore, it is important to note that the system size may play a pivotal role by the extinction of a strategy even if the latter is very large. Such undesirable system-size effects were observed on interaction graphs hosting up to $N=2.5 \cdot 10^7$ players. To eschew these artefact, we introduced a tiny mutation during the simulations; in particular, after every MCS the strategy of $N\cdot 10^{-5}$ players was changed randomly. This small mutation rate does not influence the positions of the phase transitions but solely ensures computationally manageable conditions in all phases.

\begin{figure}
\centerline{\epsfig{file=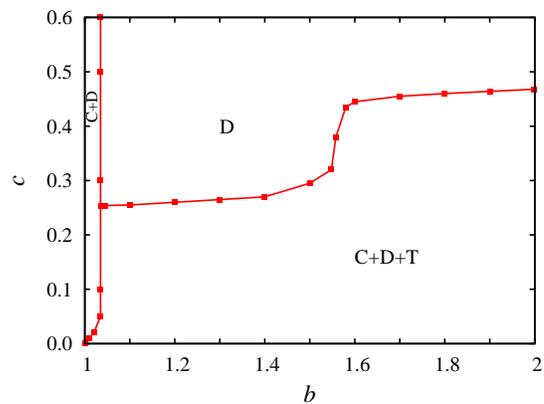,width=8cm}}
\caption{(color online) Full $b-c$ phase diagram for the three-strategy evolutionary game on the square lattice obtained via Monte Carlo simulations (boxes; connecting lines are just to guide the eye). The notation of phases and separation lines is the same as in Fig.~\ref{phd_4p}.}
\label{sqr_mc}
\end{figure}

Figure~\ref{sqr_mc} features the phase diagram for the square lattice as obtained via Monte Carlo simulations of the three-strategy evolutionary game. Apart from the shifts of the corresponding regions the results are in good agreement with the phase diagram presented in Fig.~\ref{phd_4p}, which was obtained via the $4$-site cluster dynamical mean-field approximation. Regardless of the differences, both approaches stress that the addition of the third strategy, and with it related emergence of a closed loop of dominance between the three strategies on structured populations, strongly facilitates the cooperative behavior even by high temptations to defect if only the monitoring costs $c$ remain lower than a threshold value (\textit{i.e.} $c<0.468$ at $b=1.99$). Most notably, the survival barrier of the stationary mixed C+D+T state in Fig.~\ref{sqr_mc} shifts towards higher $c$ if $b$ is increased. This feature is in agreement with the general character of systems whose dynamics is governed by a closed loop of dominance. Namely, the direct support of a player uplifts foremost the survival probability of its `predator', rather than the density of the directly supported player itself \cite{frean_prsb01, szabo_pre00a}. In our case increasing $b$ directly support the strategy D. Accordingly, the T strategy benefits from the increase of $b$ leading to the shrinking of the pure D region, as depicted in Fig.~\ref{sqr_mc}. It is also worth noticing that the border between C+D and D phases becomes independent of $c$. This is because the game becomes identical to the prisoner's dilemma game when T die out. In this case the extinction of cooperators depends only on $b$.

\begin{figure}
\centerline{\epsfig{file=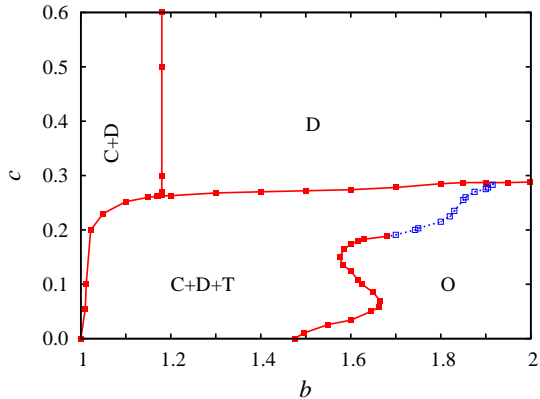,width=8cm}}
\caption{(color online) Full $b-c$ phase diagram for the three-strategy evolutionary game on the random regular graph obtained via Monte Carlo simulations. Closed (open) boxes connected by red solid (blue dashed) lines denote continuous second order (discontinuous first order) phase transitions between marked phases. The notation of phases is the same as in Fig.~\ref{phd_2p}.}
\label{rrg_mc}
\end{figure}

\begin{figure}
\centerline{\epsfig{file=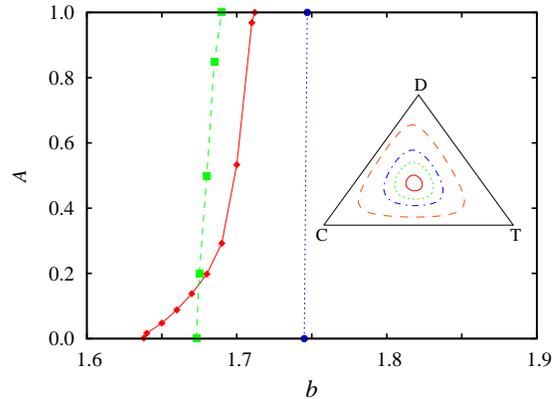,width=8cm}}
\caption{(Color online) Area of orbits $A$ in the ternary diagram in dependence on $b$ for three different values of the cost $c$. The phase transition is continuous at $c=0.1$ (red diamonds), but becomes discontinuous at $c=0.185$ (green squares). When increasing $c$ further the jumps become even sharper, as depicted by blue circles obtained at $c=0.2$. The inset shows the orbits in the ternary diagram for $b=1.64, 1.66, 1.68,$ and $1.70$ (from center) when $c=0.1$.}
\label{jump}
\end{figure}

Further supplementing the above-presented findings of the $k$-site cluster dynamical mean-field approximations, we show in Fig.~\ref{rrg_mc} the $b-c$ phase diagram obtained via Monte Carlo simulations of the evolutionary game on the random regular graph. Presented results clearly support our previous observation regarding the impact of the random interaction topology in that the oscillatory phase (O) appears when the temptation to defect $b$ is high enough, and moreover, that this phase may emerge very suddenly via a discontinuous phase transition. The nature of these oscillations on the random regular graph can be characterized succinctly by the area $A$ inside each corresponding orbit in the ternary diagram \cite{szabo_jpa04}. In particular, to demonstrate the different characters of the phase transitions from stationary C+D+T to the oscillating phases at different values of the cost $c$, we plot $A$ in dependence on $b$ for three different values of $c$ in Fig.~\ref{jump}. By small costs the oscillating phase emerges gradually when $b$ is enlarged (red diamonds in Fig.~\ref{jump}). As the ternary diagram nested in Fig.~\ref{jump} shows, the orbits become larger and larger as $b$ increases, until eventually the borders of the triangle are reached. By such high values of $b$ the three strategies cyclically exchange almost complete dominance sequentially (C $\rightarrow$ T $\rightarrow$ D $\rightarrow$ C) in an oscillatory fashion, qualitatively similar as depicted above in Fig.~\ref{rare}. Accordingly, the phase transition at $c=0.1$ in Fig.~\ref{jump} can be described by a continuous $A(b)$ function. By larger costs, however, a finite jump in the order parameter $A$ signals that the transition becomes discontinuous (green squares in Fig.~\ref{jump}), \textit{i.e.} when the system enters the O phase the smallest orbit still has a finite area. This phenomenon is further amplified at even higher values of $c$, where the transition becomes so sharp that the system starts oscillating with the maximal amplitude instantaneously (blue circles in Fig.~\ref{jump}). This behavior is in full agreement with the predictions of the $2$-site cluster dynamical mean-field approximation.

\begin{figure}
\centerline{\epsfig{file=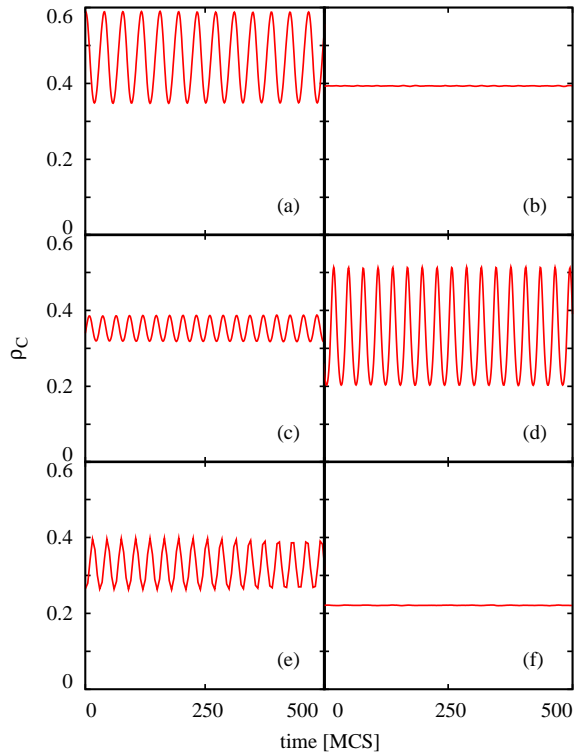,width=8cm}}
\caption{(Color online) Time evolution of the density of cooperators $\rho_{\rm C}$ for different values of $c$, as obtained via Monte Carlo simulations of the three-strategy evolutionary game on the random regular graph of $N=2.5 \cdot 10^7$ players with $b=1.6$ being fixed. Values of $c$ are: $0.01$ (a), $0.08$ (b), $0.125$ (c), $0.15$ (d), $0.17$ (e), and $0.25$ (f). Relaxation times have been discarded, hence the starting time in all panels is arbitrary. Note also that the same scale is used in all panels to ensure better comparison options.}
\label{Q1b60}
\end{figure}

Finally, the rich variety of dynamical states that can be observed within the examined three-strategy evolutionary game containing a closed loop of dominance can be demonstrated if we change the value of the cost $c$ at fixed value of the temptation to defect $b$. This also highlights the remarkable sensitivity of cyclical interactions on the parameters (in this case $c$) that define the strength of dominance between the strategies (species). As it is shown in Fig.~\ref{Q1b60}, the system is in an oscillatory phase at small $c$ values (panel a). If we increase the cost, the system arrives at a stationary C+D+T phase that is characterized by time-independent densities of the three strategies (panel b). Increasing $c$ further, the three densities start oscillating again, whereby the amplitude of these oscillations increases with larger values of $c$ (panels c, d and e). Increasing $c$ yet again, the oscillations disappear anew (panel f). At even larger values of $c$ the strategies C and T die out, leaving the game trapped in an absorbing D state (not shown). Summarizing briefly, the smooth variation of parameters can result in relevant changes of the intensity of dominance between the three strategies, which manifest as differently classified behaviors in the corresponding phase diagram.

Before summarizing, we would like to mention that the Monte Carlo simulations of the considered evolutionary game on the random regular graph indicate a transition from the O to the D phase that is in close agreement with the $2$-site cluster dynamical mean-field approximation results presented in Fig.~\ref{rare}. Namely, it can be shown that these phase transitions are continuous provided we choose as the order parameter the fraction of the time during which the system is not in the almost complete defection state.
Finally, we should stress that our observations are not valid only for the weak prisoner's dilemma parametrization. For example, using $S=-0.05$ while keeping $R=1$ and $P=0$, the topology of the $b-c$ phase diagram on random regular graphs is identical to the one plotted in Fig.~\ref{rrg_mc}. Naturally, quantitative differences are present, \textit{e.g.} the tricritical point where D, C+D and C+D+T phases meet is located at $b=1.067$ and $c=0.269$.

\section{Summary}

We have studied an evolutionary game with three strategies on the square lattice and the random regular graph via dynamical mean-field approximations and Monte Carlo simulations. Both approaches support that the introduction of the (third) tit-for-tat strategy in structured population helps to maintain the `diversity' of strategies despite of the absence of strict cyclic dominance, as defined by the payoff matrix. The coexistence of strategies prevails irrespective of the interaction structure, thus supporting the survival of cooperators even at the highest temptations to defect, provided the monitoring costs remain reasonably low. The cost of executing the tit-for-tat strategy introduces a powerful parameter to the evolutionary game, enabling us to tune the strength of effective cyclic dominance among the three participating strategies on structured populations. As a result, a rich variety of phases and phase transitions can be observed despite of the minimal number of parameters.

When the square lattice interaction topology is replaced by the random regular graph, a new oscillating phase emerges. Thus, the random topology has a similar impact on the three-strategy evolutionary game as was reported before for the voluntary PD games \cite{szabo_pre04b} and for the rock-paper-scissors games in the presence of diffusion or long range connections \cite{szabo_jpa04, reichenbach_n07, efimov_pre08}. Here the phase transitions from the stationary to the oscillating phase can be continuous (second-order) or discontinuous (first-order). The discontinuous transitions are particularly fascinating, in that just a tiny change in parameter values can result in full-blown oscillations having maximal amplitude from a state with constant strategy densities. Another noteworthy property is the extreme sensitivity of evolutionary games incorporating a closed loop of dominance, where variations of a single parameter may evoke recursive oscillatory $\rightarrow$ stationary $\rightarrow$ oscillatory phase transitions. These features highlight the diversity of feasible dynamical states and associated phase transition in evolutionary games with tunable cyclical interactions, which not only facilitate cooperation through enhanced `diversity' of strategies, but also provide interesting results related to their dynamics.

\begin{acknowledgments}
Authors acknowledge support from the Hungarian National Research Fund (grant K-73449), the Bolyai Research Scholarship, the Slovenian Research Agency (grant Z1-2032), and the Slovene-Hungarian bilateral incentive (grant BI-HU/09-10-001).
\end{acknowledgments}


\begin{thebibliography}{99}

\bibitem{sinervo_n96}
B. Sinervo and C. M. Lively, Nature (London) \textbf{380}, 240 (1996).

\bibitem{kerr_n02}
B. Kerr, M. A. Riley, M. W. Feldman, and B. J. M. Bohannan, Nature (London) \textbf{418}, 171 (2002).

\bibitem{burrows_mep98}
M. T. Burrows and S. J. Hawkins, Mar. Ecol. Prog. \textbf{167}, 1 (1998).

\bibitem{hofbauer_98}
J. Hofbauer and K. Sigmund, \textit{Evolutionnary Games and Population Dynamics} (Cambridge University Press, Cambridge, 1998).

\bibitem{frean_prsb01}
M. Frean and E. R. Abraham, Proc. R. Soc. Lond. B \textbf{268}, 1323 (2001);
O. Malcai, O. Biham, P. Richmond, and S. Solomon, Phys. Rev. E \textbf{66}, 031102 (2002).

\bibitem{szolnoki_pre04}
A. Szolnoki and G. Szab{\'o}, Phys. Rev. E \textbf{70}, 037102 (2004).

\bibitem{reichenbach_prl07}
T. Reichenbach, M. Mobilia, and E. Frey, Phys. Rev. Lett. \textbf{99}, 238105 (2007);
M. Peltom{\"a}ki and M. Alava, Phys. Rev. E \textbf{78}, 031906 (2008).

\bibitem{efimov_pre08}
A. Efimov, A. Shabunin, and A. Provata, Phys. Rev. E \textbf{78}, 056201 (2008).

\bibitem{czaran_pnas02}
T. L. Cz{\'a}r{\'a}n, R. F. Hoekstra, and L. Pagie, Proc. Natl. Acad. Sci. USA \textbf{99}, 786 (2002).

\bibitem{hauert_s02}
C. Hauert, S. De~Monte, J. Hofbauer, and K. Sigmund, Science \textbf{296}, 1129 (2002).

\bibitem{maynard_n73}
J. Maynard~Smith and G. R. Price, Nature (London) \textbf{246}, 15 (1973).

\bibitem{rasmussen_s04}
S. Rasmussen, L. Chen, D. Deamer, D. Krakauer, N. Packard, P. Stadler, and M. Bedau, Science \textbf{303}, 963 (2004).

\bibitem{maynardsmith_82}
J. Maynard~Smith, \textit{Evolution and the Theory of Games} (Cambridge University Press, Cambrdige, 1982).

\bibitem{nowak_06}
M. A. Nowak, \textit{Evolutionary Dynamics} (Harvard University Press, Harvard, 2006).

\bibitem{szabo_jpa04}
G. Szab\'{o}, A. Szolnoki, and R. Izs\'{a}k, J. Phys. A \textbf{37}, 2599 (2004).

\bibitem{szabo_pr07}
G. Szab\'{o} and G. F\'{a}th, Phys. Rep. \textbf{446}, 97 (2007).

\bibitem{traulsen_09}
A. Traulsen and C. Hauert, {\it Reviews of Nonlinear Dynamics and Complexity} Vol. II. (ed. H.G. Schuster) (Wiley-VCH, Berlin, 2009).

\bibitem{szollosi_pre08}
G. J. Sz\"{o}ll\H {o}si and I. Der\'{e}nyi, Phys. Rev. E \textbf{78}, 031919 (2008).

\bibitem{reichenbach_n07}
T. Reichenbach, M. Mobilia, and E. Frey, Nature (London) \textbf{448}, 1046 (2007).

\bibitem{reichenbach_pre06}
T. Reichenbach, M. Mobilia, and E. Frey, Phys. Rev. E \textbf{74}, 051907 (2006);
M. Perc and A. Szolnoki, New J. Phys. \textbf{9}, 267 (2007);
J. C. Claussen and A. Traulsen, Phys. Rev. Lett. \textbf{100}, 058104 (2008).

\bibitem{szabo_prl02}
G. Szab\'{o} and C. Hauert, Phys. Rev. Lett. \textbf{89}, 118101 (2002).

\bibitem{semmann_n03}
D. Semmann, H.-J. Krambeck, and M. Milinski, Nature (London) \textbf{425}, 390 (2003).

\bibitem{peltomaki_pre08b}
M. Peltom{\"a}ki, M. Rost, and M. Alava, Phys. Rev. E \textbf{78}, 050903(R) (2008).

\bibitem{imhof_pnas05}
L. A. Imhof, D. Fudenberg, and M. A. Nowak, Proc. Natl. Acad. Sci. USA \textbf{102}, 10797 (2005).

\bibitem{axelrod_84}
R. Axelrod, \textit{The Evolution of Cooperation} (Basic Books, New York, 1984).

\bibitem{nowak_n92b}
M. A. Nowak and K. Sigmund, Nature (London) \textbf{355}, 250 (1992).

\bibitem{nowak_n93}
M. A. Nowak and K. Sigmund, Nature (London) \textbf{364}, 56 (1993).

\bibitem{imhof_jtb07}
L. A. Imhof, D. Fudenberg, and M. A. Nowak, J. Theor. Biol. \textbf{247}, 574 (2007).

\bibitem{nowak_n04}
M. A. Nowak, A. Sasaki, C. Taylor, and D Fudenberg, Nature (London) \textbf{428}, 646 (2004);
J. C. Claussen and A. Traulsen, Phys. Rev. E \textbf{71}, 025101(R) (2005);
A. Traulsen, J. C. Claussen, and C. Hauert, Phys. Rev. Lett. \textbf{95}, 238701 (2005).

\bibitem{macy_pnas02}
M. W. Macy and A. Flache, Proc. Natl. Acad. Sci. USA \textbf{99}, 7229 (2002);
F. C. Santos, J. M. Pacheco, and T. Lenaerts, Proc. Natl. Acad. Sci. USA \textbf{103}, 3490 (2006);
F. Fu, T. Wu, and L. Wang, Phys. Rev. E \textbf{79}, 036101 (2009);
D.-P. Yang, J. W. Shuai, H. Lin, and C.-X. Wu., Physica A \textbf{388}, 2750 (2009);
S. Van Segbroeck, F.C. Santos, T. Lenaerts, J.M. Pacheco, Phys. Rev. Lett. \textbf{102}, 058105 (2009);
X. Chen and L. Wang, Phys. Rev. E \textbf{77}, 017103 (2008);
W.-X. Wang, J. L\"{u}, G. Chen, and P. M. Hui, Phys. Rev. E \textbf{77}, 046109 (2008);
G. Szab\'{o}, A. Szolnoki, and J. Vukov, EPL \textbf{87}, 18007 (2009);
L. Wardil and J. K. L. da Silva, EPL \textbf{86}, 38001 (2009);
H.-X. Yang, W.-X. Wang, Z.-X. Wu, Y.-C. Lai, and B.-H. Wang, Phys. Rev. E \textbf{79}, 056107 (2009);
W.-B. Du, X.-B. Cao, L. Zhao, and M.-B. Hu, Physica A \textbf{388}, 4509 (2009);
D.-P. Yang, J. W. Shuai, H. Lin H, and C.-X. Wu, Physica A \textbf{388}, 2750 (2009);
F. Fu, T. Wu, and L. Wang, Phys. Rev. E \textbf{79}, 036101 (2009).

\bibitem{nowak_n92}
M. A. Nowak and R. M. May, Nature (London) \textbf{359}, 826 (1992).

\bibitem{szabo_pre00a}
G. Szab{\'o}, T. Antal, P. Szab{\'o}, and M. Droz, Phys. Rev. E \textbf{62}, 1095 (2000).

\bibitem{dickman_pre01}
R. Dickman, Phys. Rev. E \textbf{64}, 016124 (2001).

\bibitem{cyc}
M. A. Nowak and K. Sigmund, Nature (London) \textbf{355}, 250 (1992);
B. C. Kirkup and M. A. Riley, Nature (London) \textbf{428}, 412 (2004);
R. E. Lenski and M. A. Riley, Proc. Natl. Acad. Sci. USA \textbf{99}, 556 (2002).

\bibitem{szabo_pre98}
G. Szab\'{o} and C. T{\H{o}}ke, Phys. Rev. E \textbf{58}, 69 (1998).

\bibitem{doebeli_pnas97}
M. Doebeli, A. Blarer, and M. Ackermann, Proc. Natl. Acad. Sci. USA \textbf{94}, 5167 (1997).

\bibitem{dickman_pra90}
R. Dickman, Phys. Rev. A \textbf{41}, 2192 (1990).

\bibitem{szabo_pre99}
G. Szab\'{o}, M. A. Santos, and J. F. F. Mendes, Phys. Rev. E \textbf{60}, 3776 (1999).

\bibitem{reichenbach_prl08}
T. Reichenbach and E. Frey, Phys. Rev. Lett. \textbf{101}, 058102 (2008).

\bibitem{peltomaki_jsm09}
M. Peltom{\"a}ki, M. Rost, and M. Alava, J. Stat. Mech. P02042 (2009).

\bibitem{szabo_pre04b}
G. Szab\'{o} and J. Vukov, Phys. Rev. E \textbf{69}, 036107 (2004).

\end{thebibliography}
\end{document}